\documentclass[aps,prl,twocolumn,showpacs,amsmath,amssymb,superscriptaddress,floatfix,nofootinbib]{revtex4-1}

\usepackage{color}         
\usepackage{graphics}      
\usepackage[pdftex]{graphicx}      
\usepackage{bm}
\usepackage{natbib}            
\usepackage[colorlinks,plainpages=false,linkcolor=blue,urlcolor=blue,citecolor=blue,pdfpagemode=UseNone,pdfstartview=FitBH]{hyperref}
\usepackage{float}

\begin{document}
\title{Digital modulation of the nickel valence state in a cuprate-nickelate heterostructure}

\author{F.~Wrobel}
\affiliation{Max Planck Institute for Solid State Research, Heisenbergstr. 1, 70569 Stuttgart, Germany}

\author{B. Geisler}
\affiliation{Department of Physics and Center for Nanointegration (CENIDE), Universit\"at Duisburg-Essen, Lotharstr. 1, 47057 Duisburg, Germany}

\author{Y.~Wang}
\affiliation{Max Planck Institute for Solid State Research, Heisenbergstr. 1, 70569 Stuttgart, Germany}

\author{G.~Christiani}
\affiliation{Max Planck Institute for Solid State Research, Heisenbergstr. 1, 70569 Stuttgart, Germany}

\author{G.~Logvenov}
\affiliation{Max Planck Institute for Solid State Research, Heisenbergstr. 1, 70569 Stuttgart, Germany}

\author{M. Bluschke}
\affiliation{Max Planck Institute for Solid State Research, Heisenbergstr. 1, 70569 Stuttgart, Germany}
\affiliation{Helmholtz-Zentrum Berlin f\"{u}r Materialien und Energie, Wilhelm-Conrad-R\"{o}ntgen-Campus BESSY II, Albert-Einstein-Str. 15, D-12489 Berlin, Germany}

\author{E. Schierle}
\affiliation{Helmholtz-Zentrum Berlin f\"{u}r Materialien und Energie, Wilhelm-Conrad-R\"{o}ntgen-Campus BESSY II, Albert-Einstein-Str. 15, D-12489 Berlin, Germany}

\author{P.~A.~van~Aken}
\affiliation{Max Planck Institute for Solid State Research, Heisenbergstr. 1, 70569 Stuttgart, Germany}

\author{R.~Pentcheva}
\email[]{rossitza.pentcheva@uni-due.de}
\affiliation{Department of Physics and Center for Nanointegration (CENIDE), Universit\"at Duisburg-Essen, Lotharstr. 1, 47057 Duisburg, Germany}

\author{E.~Benckiser}
\email[]{E.Benckiser@fkf.mpg.de}
\affiliation{Max Planck Institute for Solid State Research, Heisenbergstr. 1, 70569 Stuttgart, Germany}

\author{B.~Keimer}
\email[]{B.Keimer@fkf.mpg.de}
\affiliation{Max Planck Institute for Solid State Research, Heisenbergstr. 1, 70569 Stuttgart, Germany}

\pacs{81.15.Hi,73.21.-b, 78.70.Dm, 71.15.Mb, 68.37.Og}


\begin{abstract}
Layer-by-layer oxide molecular beam epitaxy has been used to synthesize cuprate-nickelate multilayer structures of composition (La$_2$CuO$_4$)$_m$/LaO/(LaNiO$_3$)$_n$. In a combined experimental and theoretical study, we show that these structures allow a clean separation of dopant and doped layers.
Specifically, the LaO layer separating cuprate and nickelate blocks provides an additional charge that, according to density functional theory calculations, is predominantly accommodated in the interfacial nickelate layers. This is reflected in an elongation of bond distances and changes in valence state, as observed by scanning transmission electron microscopy and  x-ray absorption spectroscopy. Moreover, the predicted charge disproportionation in the nickelate interface layers leads to a thickness-dependent metal-to-insulator transition for $n=2$, as observed in electrical transport measurements. The results exemplify the perspectives of charge transfer in metal-oxide multilayers to induce doping without introducing chemical and structural disorder.
\end{abstract}

\maketitle

The detailed understanding of interfacial structures and electronic energy levels in semiconductor heterostructures synthesized by molecular beam epitaxy (MBE) has driven many advances in research and technology \cite{Kroemer2001}. The last decade has seen tremendous progress in the application of MBE and other layer deposition methods to metal-oxide compounds that exhibit quantum many-body phenomena such as correlation-driven metal-insulator transitions, complex magnetism, multiferroicity, and unconventional superconductivity \cite{Hwang2012,Martin2012,Stemmer2014,Chen2017}. These developments open up key new perspectives for the management of disorder generated by doping. Bulk metal-oxides can be doped by chemical substitution, \cite{Attfield2001} but the structural and chemical disorder generated in this way can drastically modify the electronic phase behavior \cite{Mathieu2006} and/or generate mesoscopic electronic inhomogeneities \cite{Dagotto2005}. Following design principles gleaned from semiconductor physics, layer-by-layer MBE deposition can mitigate these complications by separating the dopants from the electronically active layers. In this way, the electron mobility in field-effect devices composed of binary and ternary metal oxides has steadily increased over the past decade \cite{Hwang2012,Stemmer2014}.

Modern oxide MBE technology also allows the layer-by-layer synthesis of periodic transition metal-oxide structures that have no bulk analog \cite{Hwang2012,Martin2012}. Scaling up the disorder management methods established for field-effect devices to complex multilayer structures has proven difficult, because electrostatic forces in conjunction with thermal diffusion of the dopant atoms at the temperatures required for atomic layer deposition impose severe limitations on the ability to create sharp interfaces between dopant and doped layers \cite{Nakagawa2006,Baiutti2015}. However, the transfer of charge across metal-oxide interfaces opens up new perspectives for modulating the valence state of transition metal oxides with minimal disorder \cite{Chen2017}. Here, we demonstrate the perspectives of this approach in the perovskite structure, which is adopted by most metal-oxides of topical interest. Perovskites are composed of \textit{A}-cations that occupy the voids of a three-dimensional network of corner-sharing metal-oxygen \textit{B}O$_6$ octahedra. We take advantage of the ability of MBE to deposit a single $A$O$^+$ layer at the interface between two perovskites with the same $A$-cation sublattice. The additional atomic layer acts as an electron donor, while at the same time preventing diffusion and intermixing of transition metal species during the growth process. In this way, the valence state of the constituent transition metal ions can be modified without creating the random disorder associated with doping of bulk perovskites via \textit{A}-site substitution or variation of the oxygen stoichiometry \cite{Mathieu2006,Dagotto2005}.

\begin{figure}[tb]
	\includegraphics[width=\columnwidth]{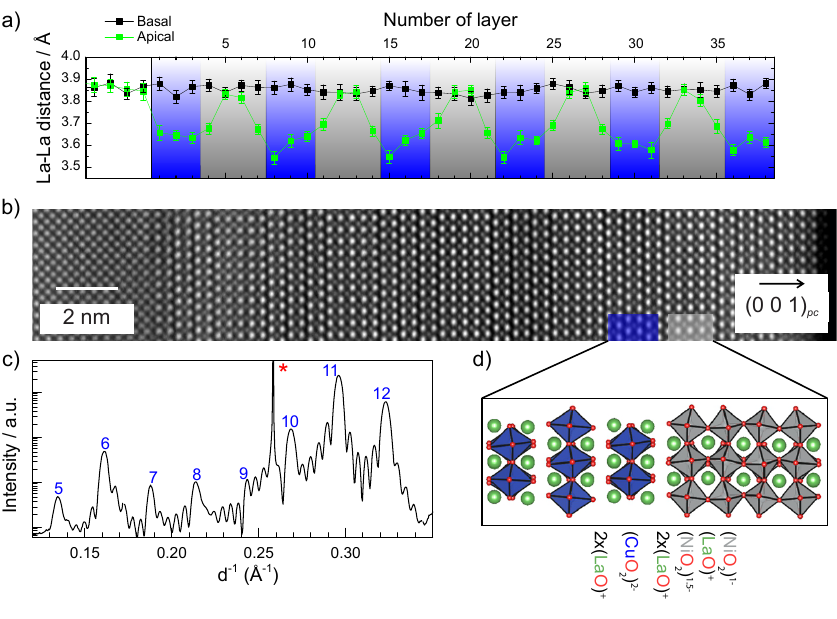}
\caption{Structural details of (LCO)$_3$/LaO/(LNO)$_4$. (a) Basal (in-plane, perpendicular to the growth direction) and apical (out-of-plane, parallel to the growth direction) distances between La atoms neighboring a transition metal site. (b) HAADF-STEM image. The black arrow indicates the growth direction (parallel to (0 0 1) in pseudo-cubic (pc) notation. (c) XRD scan along the (0 0 l)$_{\rm pc}$ direction. The substrate peak is marked with a red asterisk. The blue numbers indicate the $l$ indices of the heterostructure and result in a $c$ axis parameter of 37.24 $\pm$ 0.01\AA. (d) Side view of the relaxed structure of the column marked in (b) from DFT+$U$ calculations. The labels indicate the individual, atomic layers of the structure and their nominal charges. Note that the two different nickelate layers have different charges.}
\label{fig1}
\end{figure}

We have realized this concept by MBE-growth of a hybrid structure of the strongly correlated metal LaNiO$_3$ (LNO), which has been the focus of much recent work on ``orbital engineering'' in oxide heterostructures \cite{Chaloupka2008,Wu2015,Fabbris2016}, and the Mott insulator La$_2$CuO$_4$ (LCO), the parent compound of a well known family of high-temperature superconductors. Inserting an additional (LaO)$^+$ layer between the perovskite blocks generates superlattices of composition [(LCO)$_m$/LaO/(LNO)$_n$]$_l$ with $m, n, l$ integer (Fig.~\ref{fig1}). By combining a comprehensive set of experimental data (including transmission electron microscopy, polarized x-ray absorption, and transport measurements) with density functional calculations, we obtain a detailed, quantitative description of the atomic positions and electronic structure of the hybrid compound. We demonstrate that the electrons donated by the LaO layer are exclusively accommodated in the LNO layers, and that they induce an interfacial charge disproportionation that strongly influences the transport properties of the hybrid structure.

Atomic layers of La, Cu, and Ni were deposited on the atomically flat (001) surface of the cubic perovskite (LaAlO$_3$)$_{0.3}$(Sr$_2$AlTaO$_6$)$_{0.7}$ (LSAT) with a lattice parameter of 3.868 \AA~ at a temperature of $600\ {}^{\circ}$C in a $3 \times 10^{-5}$ mbar ozone atmosphere.  A detailed description of the oxide MBE system can be found in Ref.~\cite{Baiutti2014}. We characterized the resulting structures by x-ray diffraction (XRD), scanning transmission electron microscopy (STEM), x-ray absorption spectroscopy (XAS), and x-ray linear dichroism (XLD). High-resolution XRD measurements were performed at the MPI beamline of the ANKA light source of the Karlsruher Institute of Technology in Germany, using photon energies of 10~keV. STEM specimens were prepared by a standard procedure, and STEM investigations were performed using a JEOL JEMARM 200CF scanning transmission electron microscope. Transport measurements were performed in van-der-Pauw geometry with Pt contacts sputtered onto the corners of the square-shaped samples. XAS measurements were carried out in the XUV diffractometer at the UE46-PGM1 soft x-ray beam line at Helmholtz-Zentrum Berlin, Germany. All XAS spectra were measured in total electron yield mode with $\sigma$- and $\pi$-polarized light at an angle $\theta = 30^{\circ}$ between the sample and the incoming light. For $\sigma$-polarization, the measured intensity $I_{\sigma}(E)$ corresponds to $I_x(E) = I_y(E)$ with the electric field in the sample plane, i.e., probing the $d_{x^2-y^2}$ orbitals. The intensity measured with $\pi$-polarized light, $I_{\pi}(E)$, needs to be corrected to obtain $I_z(E) = \frac{4}{3}I_{\pi}(E) - \frac{1}{3}I_x(E)$, for the electric field perpendicular to the sample plane, i.e., probing the $d_{3z^2-r^2}$ orbitals.

\begin{figure}[tb]
	\center\includegraphics[width=0.8\columnwidth]{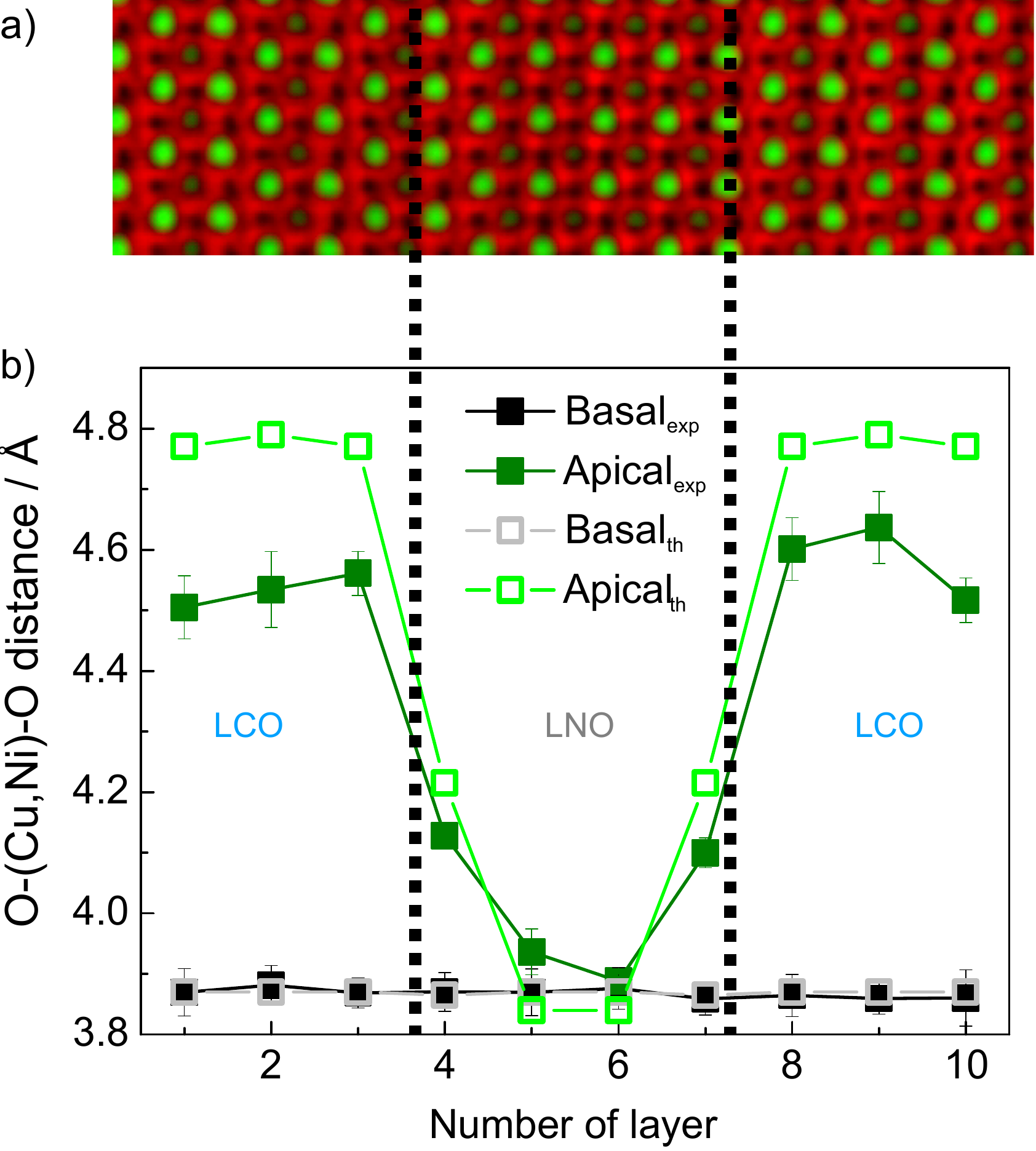}
\caption{(a) Superimposed annular bright field (ABF, red) and high-angle annular dark field (HAADF, green) images of the sample shown in Figure~\ref{fig1}. While ABF mode is sensitive to light atoms, HAADF mode is sensitive to heavy atoms. La atoms appear bright green, Cu and Ni atoms dark green, and O atoms black. The black dashed lines serve as guide to the eye to distinguish cuprate from nickelate layers. The corresponding measured O-(Cu,Ni)-O distances (index: exp) for basal and apical O atoms are shown in (b) together with the calculated ones (index: th).}
\label{fig2}
\end{figure}

Density functional theory (DFT$+U$) calculations further elucidate the structural and electronic properties. First-principles simulations of $m = 2, 3$ and $n = 2, 4$ structures were performed in the framework of spin-polarized DFT by using the Quantum Espresso code~\cite{Giannozzi2009} with the PBE exchange-correlation functional~\cite{Perdew1996} on a mesh of $12\times 12\times 2$ $k$ points. We considered $\sqrt{2}a \times \sqrt{2}a \times 2c$ supercells, rotated by 45$^{\circ}$ about the [001] axis with respect to the (pseudo-)cubic perovskite unit cell, and fixed the in-plane lattice parameter to the LSAT substrate value 3.868 \AA. On-site static electronic correlation effects~\cite{Cococcioni2005,Anisimov1993} were taken into account using $U = 4$ eV for Ni $3d$ and Cu $3d$ orbitals. Wave functions and density were expanded into plane waves up to cutoff energies of 35 and 350~Ry, respectively. The results for structures with $m = 2, 3$ and $n = 4$, referred to as 2/4 and 3/4, respectively, are very similar. Therefore we focus our discussion on the 3/4 structure in the following.

\begin{figure}[tb]
	\center\includegraphics[width=0.9\columnwidth]{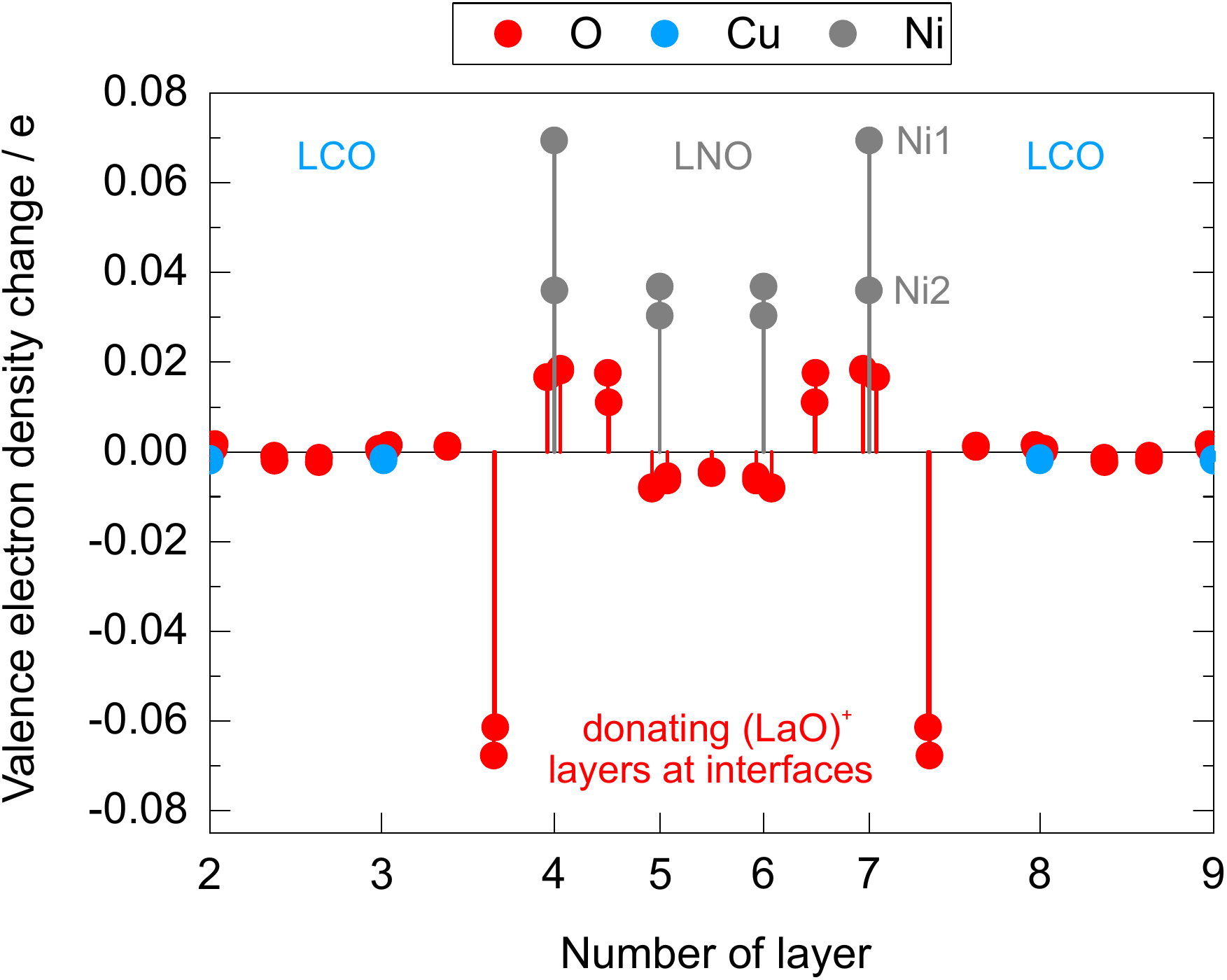}
\caption{DFT+$U$ results for the local valence electron differences relative to bulk as function of the ion position in [001] direction of 3/4. Each layer contains two inequivalent transition metal ions. The split Ni1 $\&$ Ni2 values reflect the charge disproportionation.}
\label{fig3}
\end{figure}

\begin{figure}[tb]
	\includegraphics[width=1.0\columnwidth]{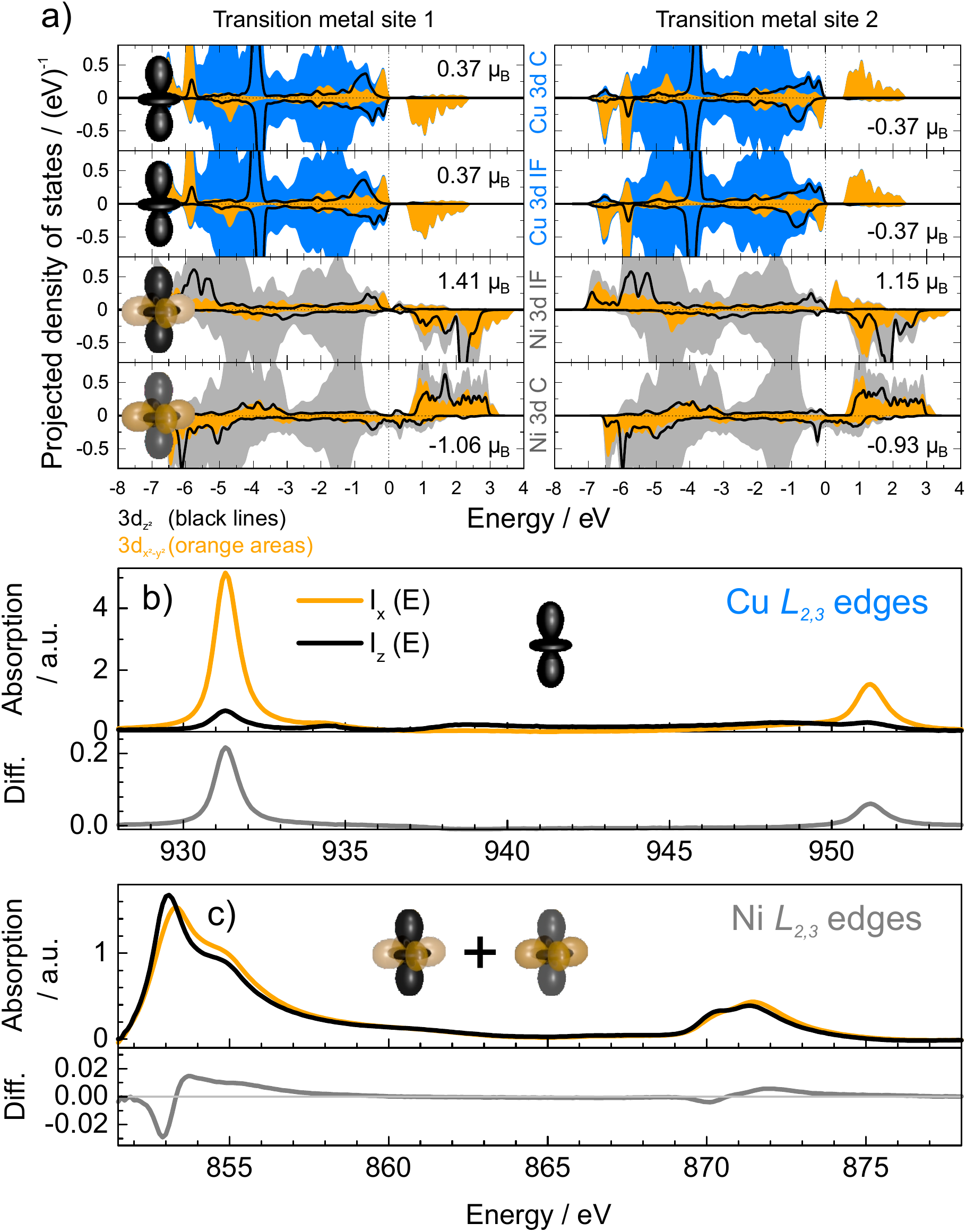}
\caption{(a) Layer-, spin-, and site-resolved calculated electronic density of states of 3/4. Only inequivalent layers are shown. Blue and grey areas depict the Cu and Ni $3d$ states in total, whereas orange areas and black lines depict the projections onto the $d_{x^2-y^2}$ and $d_{3z^2-r^2}$ orbitals. The two columns represent the two inequivalent transition metal sites per layer. The printed numbers are the local magnetic moments. The images of the $d$ orbitals represent the number and anisotropy of electrons. (b) and (c) XAS spectra measured with the electric field parallel to the sample surface (orange and black lines) together with the normalized dichroic signal (grey line) defined by $(I_x (E)-I_z (E))/((2I_x + I_z)/3)$ at the Cu (b) and Ni (c) $L$ edges.}
\label{fig4}
\end{figure}

Figure~\ref{fig1} displays STEM and XRD data of a representative 3/4 sample, which demonstrate that the interfaces are atomically sharp, and that the superlattice periodicity is maintained over the entire multilayer structure. Figure~\ref{fig2} compares experimental and theoretical results for the O-(Cu,Ni)-O distances that are of key importance for the electronic properties. We distinguish between basal (in-plane, black) and apical (out-of-plane, green) O atoms. The experimental bond distances were extracted from the STEM image in Fig.~\ref{fig2}(a). The theoretical values were obtained by DFT$+U$ calculations in which the basal distances were set to 3.87 \AA~(due to the coherent epitaxial growth on the LSAT substrate), and the apical distances were free to vary across the different layers. The trend of the bond distances found in the experiment is well reproduced by the DFT calculations.

An important structural feature is the elongation of the interfacial NiO$_6$ octahedra in the [001] direction (layers 4 and 7 in Fig.~\ref{fig2}(b)). From STEM (DFT) we obtain a ratio of apical/basal O-Ni-O distances of about 1.06 (1.08) for interfacial octahedra and about 1.01 (1.00) for the inner octahedra. In contrast, the CuO$_6$ octahedra are identical in size within the error bar, with an apical O-Cu-O distance of 4.6 \AA ~(4.8 \AA). The DFT bulk value is 4.9 \AA.

To gain direct insight into the electronic structure, we performed DFT$+U$ calculations and polarization-dependent XAS measurements. In the following, we discuss the electronic and orbital reconstruction, as well as electronic transport.

\begin{table}[b]
\centering
\caption{DFT$+U$ results for the 2/4 and 3/4 hybrid structures. We considered antiferromagnetic (AFM), ferromagnetic (FM), and nonmagnetic (NM) order in the LNO layers. We use $\sqrt{2} a \times \sqrt{2} a \times 2c$ supercells, setting $a$ to the substrate lattice parameter, $a = 3.868$ \AA . $c$ was either set to the value measured by XRD ($c_{XRD}$) or optimized ($c_{opt}$) for FM spin alignment in LNO. The total energy difference $\Delta E$ with respect to the AFM state, the total magnetic moment $M$ per supercell, the ratio of unoccupied Ni $e_g$ states $X$ ($d_{x^2-y^2}$ /$d_{3z^2-r^2}$) for interfacial (IF) and central (C) layers, and the occurrence of charge disproportionation (CD) are given for each configuration.}
\label{table}
\begin{tabular}{lccccl}
\hline
  & \textbf{\textit{c} (\AA )} & \textbf{$\Delta E$ (meV/cell)} & \textbf{$M$ ($\mu_B$/cell)} & \textbf{$X_{\rm IF / C}$} & \textbf{CD} \\
\hline
\multicolumn{2}{l}{\textit{Stacking: 2/4}}                                 &                                         &                                            &                                                         &             \\
AFM            & \begin{tabular}[c]{@{}c@{}}30.90\\  ($c_{XRD}$)\end{tabular}   & 0                                       & 4                                          & 0.84 / 0.91                                             & yes         \\
AFM            & \begin{tabular}[c]{@{}c@{}}31.36   \\ ($c_{opt}$)\end{tabular} & 0                                       & 4                                          & 0.85 / 0.92                                             & yes         \\
FM             & \begin{tabular}[c]{@{}c@{}}30.90 \\ ($c_{XRD}$)\end{tabular}   & 59                                      & 20                                         & 0.92 / 1.00                                             & no          \\
FM             & \begin{tabular}[c]{@{}c@{}}31.36 \\ ($c_{opt}$)\end{tabular}   & 105                                     & 20                                         & 0.93 / 1.01                                             & no          \\
NM             & \begin{tabular}[c]{@{}c@{}}30.90 \\ ($c_{XRD}$)\end{tabular}   & 4563                                    & 0                                          & 0.88 / 0.94                                             & no          \\
\multicolumn{2}{l}{\textit{Stacking: 3/4}}                                 &                                         &                                            &                                                         &             \\
AFM            & \begin{tabular}[c]{@{}c@{}}37.24\\   ($c_{XRD}$)\end{tabular}  & 0                                       & 4                                          &0.84 / 0.91 & yes         \\
AFM            & \begin{tabular}[c]{@{}c@{}}38.05\\   ($c_{opt}$)\end{tabular}  & 0                                       & 4                                          & 0.85 / 0.92 & yes         \\
FM             & \begin{tabular}[c]{@{}c@{}}37.24\\   ($c_{XRD}$)\end{tabular}  & 1                                       & 20                                         & 0.92 / 1.01 & no          \\
FM             & \begin{tabular}[c]{@{}c@{}}38.05\\   ($c_{opt}$)\end{tabular}  & 19                                      & 20                                         & 0.90 / 0.99 & yes         \\
NM             & \begin{tabular}[c]{@{}c@{}}37.24\\   ($c_{XRD}$)\end{tabular}  & 4423                                    & 0                                          & 0.89 / 0.96 & no          \\ \hline
\end{tabular}
\end{table}

\begin{figure}[t]
	\center\includegraphics[width=0.90\columnwidth]{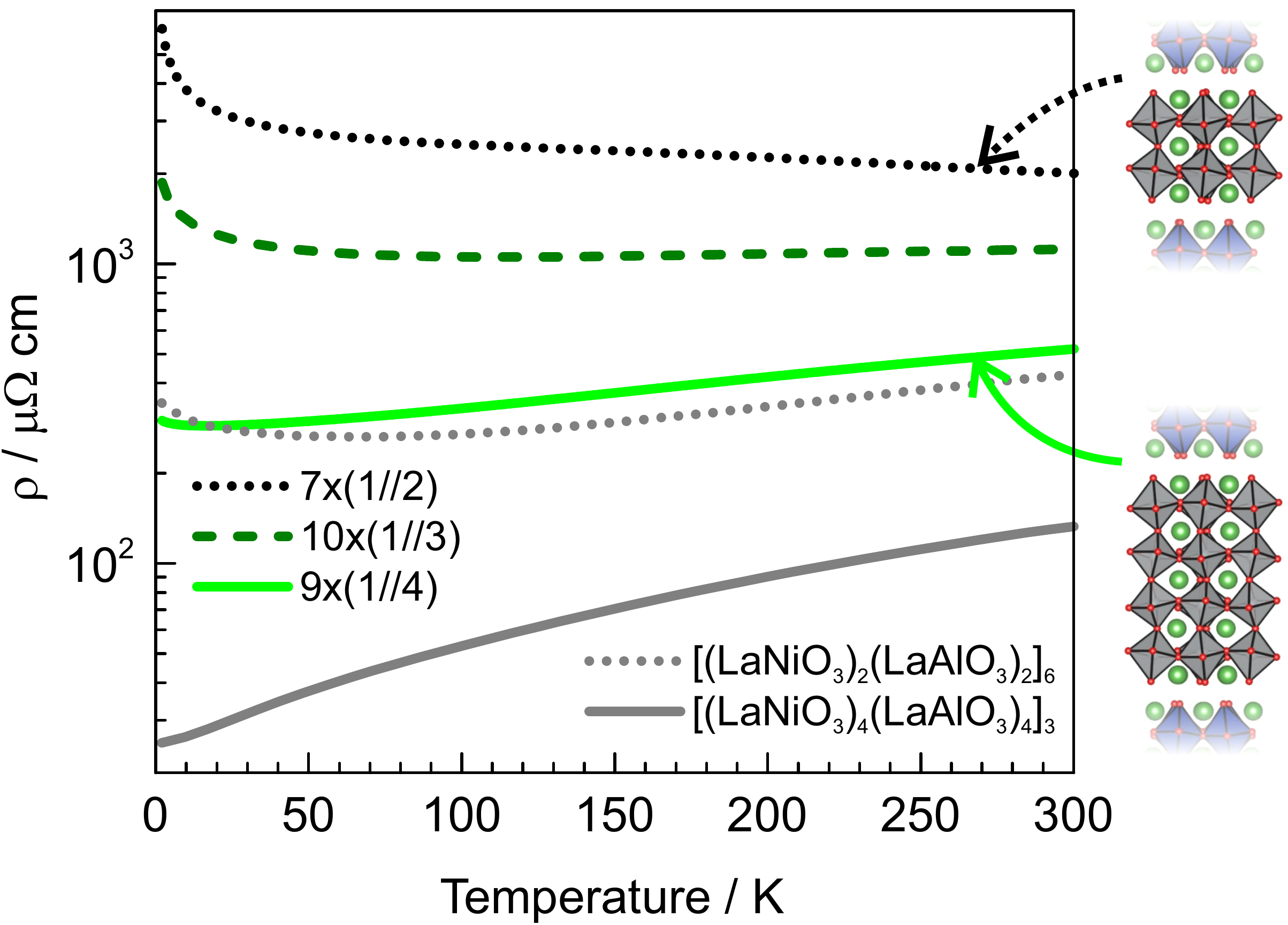}
\caption{Temperature-dependent resistivity of [(La$_2$CuO$_4$)$_m$/LaO/(LaNiO$_3$)$_n$]$_l$ ($m = 1$, $n = 2, 3, 4$ and $l = 7, 10, 9$) with average formal Ni valences of 2.5+, 2.67+, and 2.75+ for $n = 2, 3, 4$, respectively, compared to [(LaNiO$_3$)$_n$(LaAlO$_3$)$_n$]$_k$ SLs with $n = 2, 4$ and $k = 6, 3$ as indicated in the legends.}
\label{fig5}
\end{figure}

The LaO layers that separate the LCO and LNO regions each dope 0.5~$e$ per $1\times 1$ unit area. According to the DFT+$U$ calculations, this surplus charge is exclusively accommodated within LNO, predominantly in its interfacial layers, while LCO remains bulk-like, as one can infer from the local valence electron density differences compared to bulk (Fig.~\ref{fig3}) and from the projected density of states (PDOS, Fig.~\ref{fig4}(a)). A similar compensation mechanism was observed recently in (LNO)$_3$/(SrTiO$_3$)$_3$(001) heterostructures~\cite{Geisler2017}. These results are in very good agreement with our XAS Cu $L$ edge measurements (Fig.~\ref{fig4}(b)) showing all characteristics of Cu$^{2+}$~\cite{Grioni1989}. In contrast, the Ni $L$ edge spectrum shows signatures of a Ni$^{2+}$ and Ni$^{3+}$ mixture (Fig.~\ref{fig4}(c)).

The octahedral coordination splits the Cu and Ni $3d$ levels into $t_{2g}$ and $e_g$ subsets, where the $t_{2g}$ set is fully occupied. The [001] elongation of the octahedra is expected to cause a preferential occupation of the $d_{3z^2-r^2}$ orbital. We measured this anisotropy with linearly polarized x-rays, where an electric field lying in (perpendicular to) the sample plane probes holes in the $d_{x^2-y^2}$ ($d_{3z^2-r^2}$) orbital. The resulting room-temperature XLD is very pronounced at the Cu $L$ edge (Fig.\ref{fig4}(b)) and originates from a Jahn-Teller distortion that lowers the energy of the $d_{3z^2-r^2}$ orbital \cite{Nunez1995}. The same effect is found, albeit less pronounced, for the Ni $L$ edge (Fig.~\ref{fig4}(c)).
These results can be readily understood from our \textit{ab initio} calculations (Tab.~\ref{table}, Fig.~\ref{fig3} and \ref{fig4}(a)), which show that the unoccupied Cu (Ni) states in an interval up to 4 eV above the Fermi energy indeed have exclusive $d_{x^2-y^2}$ (prevailing $d_{x^2-y^2}$) character. To quantify these findings for the Ni $e_g$ orbitals, we define the ratio of unoccupied states (denoted by $h$),
\begin{equation}
X = \frac{h_{d_{3z^2 - r^2}}}{h_{d_{x^2-y^2}}}
\end{equation}
as it can be directly obtained from XLD measurements via sum rules~\cite{Laan1994}. While the DFT results show $X = 1$ for bulk LNO, we mostly find $X < 1$ for the heterostructures corresponding to a stronger averaged $d_{3z^2-r^2}$ occupation in the LNO region (see Tab.~\ref{table}). These results agree with the XLD measurements which probe the averaged Ni occupation and yield $X = 0.94$. The calculations show that the effect is most pronounced in the interface layers ($X_{\rm IF} = 0.84$), which correlates with the elongation of the NiO$_6$ octahedra in the [001] direction. The corresponding value in the central layers layers is $X_{\rm C} = 0.91$.

An interesting aspect of our DFT+$U$ results is a charge disproportionation between neighboring in-plane Ni sites (Ni1 and Ni2, Fig.~\ref{fig3}), predominantly in the interface layers, which results in different local magnetic moments as noted in Fig.~\ref{fig4}(a) (left and right panels). This effect opens a band gap in the interface layers (Fig.~\ref{fig4}(a)). Similar effects have been reported recently in DFT calculations for (LNO)$_1$/(LaAlO$_3$)$_1$(001) heterostructures~\cite{Freeland2011,Blanca2011} and bulk rare-earth nickelates~\cite{Park2012,Johnston2014}. The inequivalence of adjacent Ni sites is also reflected in a difference in NiO$_6$ octahedral volume within the interface layer of 4.3$\%$ (11.02 \AA$^3$ ~and 10.57 \AA$^3$).

\begin{figure}[t]
	\includegraphics[width=0.99\columnwidth]{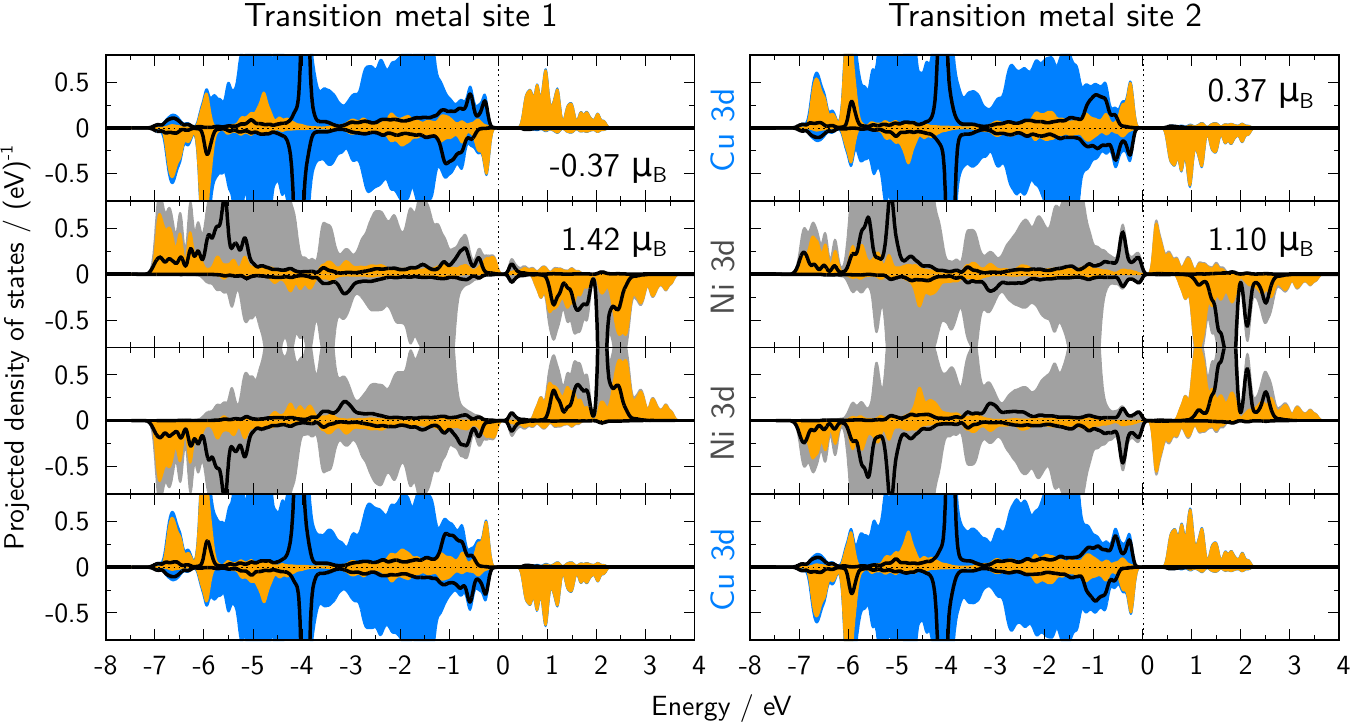}
\caption{Layer-, spin-, and site-resolved calculated electronic density of states of 3/2. Only relevant layers are shown. Blue and grey areas
depict the Cu and Ni $3d$ states in total, whereas orange areas and black lines depict the projections onto the $d_{x^2-y^2}$ and $d_{3z^2-r^2}$ orbitals. The two
columns represent the two inequivalent transition metal sites per layer. The printed numbers are the local magnetic moments. The charge
disproportionation in conjunction with the layerwise AFM order in the nickelate opens a band gap of 0.28 eV.}
\label{fig6}
\end{figure}

The presence of interfacial disproportionation is supported by the in-plane transport measurements on structures with $m = 1$ and decreasing LNO layer thickness $n = 4, 3, 2$ (Fig.~\ref{fig5}). The temperature-dependent resistivity shows metallic behavior for $n = 4$, which we attribute to currents running through the inner, at most weakly disproportionated, metallic LNO layers. When decreasing LNO from 4 to 3 monolayers we observe a metal-to-semiconductor transition, and finally for $n = 2$ a semiconducting behavior. Consistent with this observation, DFT results for $n = 2$ indicate a band gap of 0.28 eV (Fig.~\ref{fig6}). To distinguish confinement and doping effects, we compare the LNO-LCO hybrid structures with [(LNO)$_n$(LaAlO$_3$)$_n$]$_k$ ($n = 2, 4$ and $k = 6, 3$) superlattices where the nickelate layers are confined to the same thickness while remaining undoped (Ni$^{3+}$). Leaving differences in the lattice parameters of the two systems aside, it is interesting to note that the resistivity of the LNO-LAO superlattice with 2 nickelate layers is comparable to the LNO-LCO hybrid structure with 4 nickelate layers. This implies that only the inner two layers are conducting, as predicted by theory.

In summary, we showed that electrons doped in the cuprate-nickelate hybrid structures are accommodated primarily in the interfacial nickelate layers, where they induce a digital modulation of the Ni valence state and a rearrangement of the Ni $3d$-orbital occupation. We observed a metal-to-semiconductor transition with decreasing nickelate thickness, which supports the disproportionation scenario predicted by DFT calculations. It is interesting to compare our results to the Ruddlesden-Popper phases of composition La$_{n+1}$Ni$_n$O$_{3n+1}$, which have been synthesized in bulk single crystal~\cite{Zhang2017} and thin film~\cite{WuKT2013,Lee2014,Wrobel2016,Lei2017} forms. Whereas in the Ruddlesden-Popper structures one electron is shared between two neighboring LNO blocks, an additional charge of half an electron is accommodated only within a single NiO$_2$ plane due to the additional separating, insulating La$_2$CuO$_4$ blocks.

Our study demonstrates the power of charge transfer across metal-oxide inerfaces to dope a large density of charge carriers into metal-oxide planes, without generating significant disorder. Layer-by-layer deposition by MBE thus offers a versatile tool for the synthesis and investigation of new correlated-electron materials.

We thank HZB for the allocation of synchrotron radiation beamtime and Eugen Weschke for valuable support at beamline UE46/Bessy II, Berlin. The TEM sample preparation by Ute Salzberger is highly appreciated. We are grateful for the support by Peter Wochner and Shyjumon Ibrahimkutty at the MPI Beamline at ANKA-KIT, Karlsruhe. We thank the German Science Foundation (DFG) for financial support under grant No. TRR 80 projects G1, G3, and G8. Computing time was granted by the Center for Computational Sciences and Simulation of the University of Duisburg-Essen (DFG grants INST 20876/209-1 FUGG, INST 20876/243-1 FUGG).

\bibliographystyle{apsrev4-1}

\begin{thebibliography}{32}%
\makeatletter
\providecommand \@ifxundefined [1]{%
 \@ifx{#1\undefined}
}%
\providecommand \@ifnum [1]{%
 \ifnum #1\expandafter \@firstoftwo
 \else \expandafter \@secondoftwo
 \fi
}%
\providecommand \@ifx [1]{%
 \ifx #1\expandafter \@firstoftwo
 \else \expandafter \@secondoftwo
 \fi
}%
\providecommand \natexlab [1]{#1}%
\providecommand \enquote  [1]{``#1''}%
\providecommand \bibnamefont  [1]{#1}%
\providecommand \bibfnamefont [1]{#1}%
\providecommand \citenamefont [1]{#1}%
\providecommand \href@noop [0]{\@secondoftwo}%
\providecommand \href [0]{\begingroup \@sanitize@url \@href}%
\providecommand \@href[1]{\@@startlink{#1}\@@href}%
\providecommand \@@href[1]{\endgroup#1\@@endlink}%
\providecommand \@sanitize@url [0]{\catcode `\\12\catcode `\$12\catcode
  `\&12\catcode `\#12\catcode `\^12\catcode `\_12\catcode `\%12\relax}%
\providecommand \@@startlink[1]{}%
\providecommand \@@endlink[0]{}%
\providecommand \url  [0]{\begingroup\@sanitize@url \@url }%
\providecommand \@url [1]{\endgroup\@href {#1}{\urlprefix }}%
\providecommand \urlprefix  [0]{URL }%
\providecommand \Eprint [0]{\href }%
\providecommand \doibase [0]{http://dx.doi.org/}%
\providecommand \selectlanguage [0]{\@gobble}%
\providecommand \bibinfo  [0]{\@secondoftwo}%
\providecommand \bibfield  [0]{\@secondoftwo}%
\providecommand \translation [1]{[#1]}%
\providecommand \BibitemOpen [0]{}%
\providecommand \bibitemStop [0]{}%
\providecommand \bibitemNoStop [0]{.\EOS\space}%
\providecommand \EOS [0]{\spacefactor3000\relax}%
\providecommand \BibitemShut  [1]{\csname bibitem#1\endcsname}%
\let\auto@bib@innerbib\@empty
\bibitem [{\citenamefont {Kroemer}(2001)}]{Kroemer2001}%
  \BibitemOpen
  \bibfield  {author} {\bibinfo {author} {\bibfnamefont {H.}~\bibnamefont
  {Kroemer}},\ }\href {\doibase 10.1103/revmodphys.73.783} {\bibfield
  {journal} {\bibinfo  {journal} {Reviews of Modern Physics}\ }\textbf
  {\bibinfo {volume} {73}},\ \bibinfo {pages} {783} (\bibinfo {year}
  {2001})}\BibitemShut {NoStop}%
\bibitem [{\citenamefont {Hwang}\ \emph {et~al.}(2012)\citenamefont {Hwang},
  \citenamefont {Iwasa}, \citenamefont {Kawasaki}, \citenamefont {Keimer},
  \citenamefont {Nagaosa},\ and\ \citenamefont {Tokura}}]{Hwang2012}%
  \BibitemOpen
  \bibfield  {author} {\bibinfo {author} {\bibfnamefont {H.~Y.}\ \bibnamefont
  {Hwang}}, \bibinfo {author} {\bibfnamefont {Y.}~\bibnamefont {Iwasa}},
  \bibinfo {author} {\bibfnamefont {M.}~\bibnamefont {Kawasaki}}, \bibinfo
  {author} {\bibfnamefont {B.}~\bibnamefont {Keimer}}, \bibinfo {author}
  {\bibfnamefont {N.}~\bibnamefont {Nagaosa}}, \ and\ \bibinfo {author}
  {\bibfnamefont {Y.}~\bibnamefont {Tokura}},\ }\href {\doibase
  10.1038/nmat3223} {\bibfield  {journal} {\bibinfo  {journal} {Nat Mater}\
  }\textbf {\bibinfo {volume} {11}},\ \bibinfo {pages} {103} (\bibinfo {year}
  {2012})}\BibitemShut {NoStop}%
\bibitem [{\citenamefont {Martin}\ and\ \citenamefont
  {Schlom}(2012)}]{Martin2012}%
  \BibitemOpen
  \bibfield  {author} {\bibinfo {author} {\bibfnamefont {L.~W.}\ \bibnamefont
  {Martin}}\ and\ \bibinfo {author} {\bibfnamefont {D.~G.}\ \bibnamefont
  {Schlom}},\ }\href {\doibase 10.1016/j.cossms.2012.03.001} {\bibfield
  {journal} {\bibinfo  {journal} {Current Opinion in Solid State and Materials
  Science}\ }\textbf {\bibinfo {volume} {16}},\ \bibinfo {pages} {199}
  (\bibinfo {year} {2012})}\BibitemShut {NoStop}%
\bibitem [{\citenamefont {Stemmer}\ and\ \citenamefont
  {Allen}(2014)}]{Stemmer2014}%
  \BibitemOpen
  \bibfield  {author} {\bibinfo {author} {\bibfnamefont {S.}~\bibnamefont
  {Stemmer}}\ and\ \bibinfo {author} {\bibfnamefont {S.~J.}\ \bibnamefont
  {Allen}},\ }\href {\doibase 10.1146/annurev-matsci-070813-113552} {\bibfield
  {journal} {\bibinfo  {journal} {Annual Review of Materials Research}\
  }\textbf {\bibinfo {volume} {44}},\ \bibinfo {pages} {151} (\bibinfo {year}
  {2014})}\BibitemShut {NoStop}%
\bibitem [{\citenamefont {Chen}\ and\ \citenamefont {Millis}(2017)}]{Chen2017}%
  \BibitemOpen
  \bibfield  {author} {\bibinfo {author} {\bibfnamefont {H.}~\bibnamefont
  {Chen}}\ and\ \bibinfo {author} {\bibfnamefont {A.}~\bibnamefont {Millis}},\
  }\href {\doibase 10.1088/1361-648x/aa6efe} {\bibfield  {journal} {\bibinfo
  {journal} {Journal of Physics: Condensed Matter}\ }\textbf {\bibinfo {volume}
  {29}},\ \bibinfo {pages} {243001} (\bibinfo {year} {2017})}\BibitemShut
  {NoStop}%
\bibitem [{\citenamefont {Attfield}(2001)}]{Attfield2001}%
  \BibitemOpen
  \bibfield  {author} {\bibinfo {author} {\bibfnamefont {J.~P.}\ \bibnamefont
  {Attfield}},\ }\href {\doibase 10.1016/S1466-6049(01)00110-6} {\bibfield
  {journal} {\bibinfo  {journal} {International Journal of Inorganic
  Materials}\ }\textbf {\bibinfo {volume} {3}},\ \bibinfo {pages} {1147}
  (\bibinfo {year} {2001})}\BibitemShut {NoStop}%
\bibitem [{\citenamefont {Mathieu}\ \emph {et~al.}(2006)\citenamefont
  {Mathieu}, \citenamefont {Uchida}, \citenamefont {Kaneko}, \citenamefont
  {He}, \citenamefont {Yu}, \citenamefont {Kumai}, \citenamefont {Arima},
  \citenamefont {Tomioka}, \citenamefont {Asamitsu}, \citenamefont {Matsui},\
  and\ \citenamefont {Tokura}}]{Mathieu2006}%
  \BibitemOpen
  \bibfield  {author} {\bibinfo {author} {\bibfnamefont {R.}~\bibnamefont
  {Mathieu}}, \bibinfo {author} {\bibfnamefont {M.}~\bibnamefont {Uchida}},
  \bibinfo {author} {\bibfnamefont {Y.}~\bibnamefont {Kaneko}}, \bibinfo
  {author} {\bibfnamefont {J.~P.}\ \bibnamefont {He}}, \bibinfo {author}
  {\bibfnamefont {X.~Z.}\ \bibnamefont {Yu}}, \bibinfo {author} {\bibfnamefont
  {R.}~\bibnamefont {Kumai}}, \bibinfo {author} {\bibfnamefont
  {T.}~\bibnamefont {Arima}}, \bibinfo {author} {\bibfnamefont
  {Y.}~\bibnamefont {Tomioka}}, \bibinfo {author} {\bibfnamefont
  {A.}~\bibnamefont {Asamitsu}}, \bibinfo {author} {\bibfnamefont
  {Y.}~\bibnamefont {Matsui}}, \ and\ \bibinfo {author} {\bibfnamefont
  {Y.}~\bibnamefont {Tokura}},\ }\href
  {https://link.aps.org/doi/10.1103/PhysRevB.74.020404} {\bibfield  {journal}
  {\bibinfo  {journal} {Phys. Rev. B}\ }\textbf {\bibinfo {volume} {74}},\ \bibinfo {pages} {020404}
  (\bibinfo {year} {2006})}\BibitemShut {NoStop}%
\bibitem [{\citenamefont {Dagotto}(2005)}]{Dagotto2005}%
  \BibitemOpen
  \bibfield  {author} {\bibinfo {author} {\bibfnamefont {E.}~\bibnamefont
  {Dagotto}},\ }\href {\doibase 10.1126/science.1107559} {\bibfield  {journal}
  {\bibinfo  {journal} {Science}\ }\textbf {\bibinfo {volume} {309}},\ \bibinfo
  {pages} {257} (\bibinfo {year} {2005})}\BibitemShut {NoStop}%
\bibitem [{\citenamefont {Nakagawa}\ \emph {et~al.}(2006)\citenamefont
  {Nakagawa}, \citenamefont {Hwang},\ and\ \citenamefont
  {Muller}}]{Nakagawa2006}%
  \BibitemOpen
  \bibfield  {author} {\bibinfo {author} {\bibfnamefont {N.}~\bibnamefont
  {Nakagawa}}, \bibinfo {author} {\bibfnamefont {H.~Y.}\ \bibnamefont {Hwang}},
  \ and\ \bibinfo {author} {\bibfnamefont {D.~A.}\ \bibnamefont {Muller}},\
  }\href {\doibase 10.1038/nmat1569} {\bibfield  {journal} {\bibinfo  {journal}
  {Nat Mater}\ }\textbf {\bibinfo {volume} {5}},\ \bibinfo {pages} {204}
  (\bibinfo {year} {2006})}\BibitemShut {NoStop}%
\bibitem [{\citenamefont {Baiutti}\ \emph {et~al.}(2015)\citenamefont
  {Baiutti}, \citenamefont {Logvenov}, \citenamefont {Gregori}, \citenamefont
  {Cristiani}, \citenamefont {Wang}, \citenamefont {Sigle}, \citenamefont {van
  Aken},\ and\ \citenamefont {Maier}}]{Baiutti2015}%
  \BibitemOpen
  \bibfield  {author} {\bibinfo {author} {\bibfnamefont {F.}~\bibnamefont
  {Baiutti}}, \bibinfo {author} {\bibfnamefont {G.}~\bibnamefont {Logvenov}},
  \bibinfo {author} {\bibfnamefont {G.}~\bibnamefont {Gregori}}, \bibinfo
  {author} {\bibfnamefont {G.}~\bibnamefont {Cristiani}}, \bibinfo {author}
  {\bibfnamefont {Y.}~\bibnamefont {Wang}}, \bibinfo {author} {\bibfnamefont
  {W.}~\bibnamefont {Sigle}}, \bibinfo {author} {\bibfnamefont {P.~A.}\
  \bibnamefont {van Aken}}, \ and\ \bibinfo {author} {\bibfnamefont
  {J.}~\bibnamefont {Maier}},\ }\href {\doibase 10.1038/ncomms9586} {\bibfield
  {journal} {\bibinfo  {journal} {Nat Commun}\ }\textbf {\bibinfo {volume}
  {6}},\ \bibinfo {pages} {8586} (\bibinfo {year} {2015})}\BibitemShut
  {NoStop}%
\bibitem [{\citenamefont {Chaloupka}\ and\ \citenamefont
  {Khaliullin}(2008)}]{Chaloupka2008}%
  \BibitemOpen
  \bibfield  {author} {\bibinfo {author} {\bibfnamefont {J.}~\bibnamefont
  {Chaloupka}}\ and\ \bibinfo {author} {\bibfnamefont {G.}~\bibnamefont
  {Khaliullin}},\ }\href
  {http://link.aps.org/doi/10.1103/PhysRevLett.100.016404} {\bibfield
  {journal} {\bibinfo  {journal} {Phys. Rev. Lett.}\ }\textbf {\bibinfo
  {volume} {100}},\ \bibinfo {pages} {016404} (\bibinfo {year}
  {2008})}\BibitemShut {NoStop}%
\bibitem [{\citenamefont {Wu}\ \emph {et~al.}(2015)\citenamefont {Wu},
  \citenamefont {Benckiser}, \citenamefont {Audehm}, \citenamefont {Goering},
  \citenamefont {Wochner}, \citenamefont {Christiani}, \citenamefont
  {Logvenov}, \citenamefont {Habermeier},\ and\ \citenamefont
  {Keimer}}]{Wu2015}%
  \BibitemOpen
  \bibfield  {author} {\bibinfo {author} {\bibfnamefont {M.}~\bibnamefont
  {Wu}}, \bibinfo {author} {\bibfnamefont {E.}~\bibnamefont {Benckiser}},
  \bibinfo {author} {\bibfnamefont {P.}~\bibnamefont {Audehm}}, \bibinfo
  {author} {\bibfnamefont {E.}~\bibnamefont {Goering}}, \bibinfo {author}
  {\bibfnamefont {P.}~\bibnamefont {Wochner}}, \bibinfo {author} {\bibfnamefont
  {G.}~\bibnamefont {Christiani}}, \bibinfo {author} {\bibfnamefont
  {G.}~\bibnamefont {Logvenov}}, \bibinfo {author} {\bibfnamefont {H.-U.}\
  \bibnamefont {Habermeier}}, \ and\ \bibinfo {author} {\bibfnamefont
  {B.}~\bibnamefont {Keimer}},\ }\href {\doibase 10.1103/PhysRevB.91.195130}
  {\bibfield  {journal} {\bibinfo  {journal} {Phys. Rev. B}\ }\textbf {\bibinfo
  {volume} {91}},\ \bibinfo {pages} {195130} (\bibinfo {year}
  {2015})}\BibitemShut {NoStop}%
\bibitem [{\citenamefont {Fabbris}\ \emph {et~al.}(2016)\citenamefont
  {Fabbris}, \citenamefont {Meyers}, \citenamefont {Okamoto}, \citenamefont
  {Pelliciari}, \citenamefont {Disa}, \citenamefont {Huang}, \citenamefont
  {Chen}, \citenamefont {Wu}, \citenamefont {Chen}, \citenamefont
  {Ismail-Beigi}, \citenamefont {Ahn}, \citenamefont {Walker}, \citenamefont
  {Huang}, \citenamefont {Schmitt},\ and\ \citenamefont {Dean}}]{Fabbris2016}%
  \BibitemOpen
  \bibfield  {author} {\bibinfo {author} {\bibfnamefont {G.}~\bibnamefont
  {Fabbris}}, \bibinfo {author} {\bibfnamefont {D.}~\bibnamefont {Meyers}},
  \bibinfo {author} {\bibfnamefont {J.}~\bibnamefont {Okamoto}}, \bibinfo
  {author} {\bibfnamefont {J.}~\bibnamefont {Pelliciari}}, \bibinfo {author}
  {\bibfnamefont {A.~S.}\ \bibnamefont {Disa}}, \bibinfo {author}
  {\bibfnamefont {Y.}~\bibnamefont {Huang}}, \bibinfo {author} {\bibfnamefont
  {Z.-Y.}\ \bibnamefont {Chen}}, \bibinfo {author} {\bibfnamefont {W.~B.}\
  \bibnamefont {Wu}}, \bibinfo {author} {\bibfnamefont {C.~T.}\ \bibnamefont
  {Chen}}, \bibinfo {author} {\bibfnamefont {S.}~\bibnamefont {Ismail-Beigi}},
  \bibinfo {author} {\bibfnamefont {C.~H.}\ \bibnamefont {Ahn}}, \bibinfo
  {author} {\bibfnamefont {F.~J.}\ \bibnamefont {Walker}}, \bibinfo {author}
  {\bibfnamefont {D.~J.}\ \bibnamefont {Huang}}, \bibinfo {author}
  {\bibfnamefont {T.}~\bibnamefont {Schmitt}}, \ and\ \bibinfo {author}
  {\bibfnamefont {M.~P.~M.}\ \bibnamefont {Dean}},\ }\href
  {https://link.aps.org/doi/10.1103/PhysRevLett.117.147401} {\bibfield
  {journal} {\bibinfo  {journal} {Phys. Rev. Lett.}\ }\textbf {\bibinfo
  {volume} {117}},\ \bibinfo {pages} {147401} (\bibinfo {year}
  {2016})}\BibitemShut {NoStop}%
\bibitem [{\citenamefont {Baiutti}\ \emph {et~al.}(2014)\citenamefont
  {Baiutti}, \citenamefont {Christiani},\ and\ \citenamefont
  {Logvenov}}]{Baiutti2014}%
  \BibitemOpen
  \bibfield  {author} {\bibinfo {author} {\bibfnamefont {F.}~\bibnamefont
  {Baiutti}}, \bibinfo {author} {\bibfnamefont {G.}~\bibnamefont {Christiani}},
  \ and\ \bibinfo {author} {\bibfnamefont {G.}~\bibnamefont {Logvenov}},\
  }\href {\doibase 10.3762/bjnano.5.70} {\bibfield  {journal} {\bibinfo
  {journal} {Beilstein J. Nanotechnol.}\ }\textbf {\bibinfo {volume} {5}},\
  \bibinfo {pages} {596} (\bibinfo {year} {2014})}\BibitemShut {NoStop}%
\bibitem [{\citenamefont {Giannozzi}\ \emph {et~al.}(2009)\citenamefont
  {Giannozzi}, \citenamefont {Baroni}, \citenamefont {Bonini}, \citenamefont
  {Calandra}, \citenamefont {Car}, \citenamefont {Cavazzoni}, \citenamefont
  {Ceresoli}, \citenamefont {Chiarotti}, \citenamefont {Cococcioni},
  \citenamefont {Dabo}, \citenamefont {Dal~Corso}, \citenamefont
  {de~Gironcoli}, \citenamefont {Fabris}, \citenamefont {Fratesi},
  \citenamefont {Gebauer}, \citenamefont {Gerstmann}, \citenamefont
  {Gougoussis}, \citenamefont {Kokalj}, \citenamefont {Lazzeri}, \citenamefont
  {Martin-Samos}, \citenamefont {Marzari}, \citenamefont {Mauri}, \citenamefont
  {Mazzarello}, \citenamefont {Paolini}, \citenamefont {Pasquarello},
  \citenamefont {Paulatto}, \citenamefont {Sbraccia}, \citenamefont {Scandolo},
  \citenamefont {Sclauzero}, \citenamefont {Seitsonen}, \citenamefont
  {Smogunov}, \citenamefont {Umari},\ and\ \citenamefont
  {Wentzcovitch}}]{Giannozzi2009}%
  \BibitemOpen
  \bibfield  {author} {\bibinfo {author} {\bibfnamefont {P.}~\bibnamefont
  {Giannozzi}}, \bibinfo {author} {\bibfnamefont {S.}~\bibnamefont {Baroni}},
  \bibinfo {author} {\bibfnamefont {N.}~\bibnamefont {Bonini}}, \bibinfo
  {author} {\bibfnamefont {M.}~\bibnamefont {Calandra}}, \bibinfo {author}
  {\bibfnamefont {R.}~\bibnamefont {Car}}, \bibinfo {author} {\bibfnamefont
  {C.}~\bibnamefont {Cavazzoni}}, \bibinfo {author} {\bibfnamefont
  {D.}~\bibnamefont {Ceresoli}}, \bibinfo {author} {\bibfnamefont {G.~L.}\
  \bibnamefont {Chiarotti}}, \bibinfo {author} {\bibfnamefont {M.}~\bibnamefont
  {Cococcioni}}, \bibinfo {author} {\bibfnamefont {I.}~\bibnamefont {Dabo}},
  \bibinfo {author} {\bibfnamefont {A.}~\bibnamefont {Dal~Corso}}, \bibinfo
  {author} {\bibfnamefont {S.}~\bibnamefont {de~Gironcoli}}, \bibinfo {author}
  {\bibfnamefont {S.}~\bibnamefont {Fabris}}, \bibinfo {author} {\bibfnamefont
  {G.}~\bibnamefont {Fratesi}}, \bibinfo {author} {\bibfnamefont
  {R.}~\bibnamefont {Gebauer}}, \bibinfo {author} {\bibfnamefont
  {U.}~\bibnamefont {Gerstmann}}, \bibinfo {author} {\bibfnamefont
  {C.}~\bibnamefont {Gougoussis}}, \bibinfo {author} {\bibfnamefont
  {A.}~\bibnamefont {Kokalj}}, \bibinfo {author} {\bibfnamefont
  {M.}~\bibnamefont {Lazzeri}}, \bibinfo {author} {\bibfnamefont
  {L.}~\bibnamefont {Martin-Samos}}, \bibinfo {author} {\bibfnamefont
  {N.}~\bibnamefont {Marzari}}, \bibinfo {author} {\bibfnamefont
  {F.}~\bibnamefont {Mauri}}, \bibinfo {author} {\bibfnamefont
  {R.}~\bibnamefont {Mazzarello}}, \bibinfo {author} {\bibfnamefont
  {S.}~\bibnamefont {Paolini}}, \bibinfo {author} {\bibfnamefont
  {A.}~\bibnamefont {Pasquarello}}, \bibinfo {author} {\bibfnamefont
  {L.}~\bibnamefont {Paulatto}}, \bibinfo {author} {\bibfnamefont
  {C.}~\bibnamefont {Sbraccia}}, \bibinfo {author} {\bibfnamefont
  {S.}~\bibnamefont {Scandolo}}, \bibinfo {author} {\bibfnamefont
  {G.}~\bibnamefont {Sclauzero}}, \bibinfo {author} {\bibfnamefont {A.~P.}\
  \bibnamefont {Seitsonen}}, \bibinfo {author} {\bibfnamefont {A.}~\bibnamefont
  {Smogunov}}, \bibinfo {author} {\bibfnamefont {P.}~\bibnamefont {Umari}}, \
  and\ \bibinfo {author} {\bibfnamefont {R.~M.}\ \bibnamefont {Wentzcovitch}},\
  }\href {\doibase 10.1088/0953-8984/21/39/395502} {\bibfield  {journal}
  {\bibinfo  {journal} {J Phys Condens Matter}\ }\textbf {\bibinfo {volume}
  {21}},\ \bibinfo {pages} {395502} (\bibinfo {year} {2009})}\BibitemShut
  {NoStop}%
\bibitem [{\citenamefont {Perdew}\ \emph {et~al.}(1996)\citenamefont {Perdew},
  \citenamefont {Burke},\ and\ \citenamefont {Ernzerhof}}]{Perdew1996}%
  \BibitemOpen
  \bibfield  {author} {\bibinfo {author} {\bibfnamefont {J.~P.}\ \bibnamefont
  {Perdew}}, \bibinfo {author} {\bibfnamefont {K.}~\bibnamefont {Burke}}, \
  and\ \bibinfo {author} {\bibfnamefont {M.}~\bibnamefont {Ernzerhof}},\ }\href
  {http://link.aps.org/doi/10.1103/PhysRevLett.77.3865} {\bibfield  {journal}
  {\bibinfo  {journal} {Phys. Rev. Lett.}\ }\textbf {\bibinfo {volume} {77}},\
  \bibinfo {pages} {3865} (\bibinfo {year} {1996})}\BibitemShut {NoStop}%
\bibitem [{\citenamefont {Cococcioni}\ and\ \citenamefont
  {de~Gironcoli}(2005)}]{Cococcioni2005}%
  \BibitemOpen
  \bibfield  {author} {\bibinfo {author} {\bibfnamefont {M.}~\bibnamefont
  {Cococcioni}}\ and\ \bibinfo {author} {\bibfnamefont {S.}~\bibnamefont
  {de~Gironcoli}},\ }\href {http://link.aps.org/doi/10.1103/PhysRevB.71.035105}
  {\bibfield  {journal} {\bibinfo  {journal} {Phys. Rev. B}\ }\textbf {\bibinfo
  {volume} {71}},\ \bibinfo {pages} {035105} (\bibinfo {year}
  {2005})}\BibitemShut {NoStop}%
\bibitem [{\citenamefont {Anisimov}\ \emph {et~al.}(1993)\citenamefont
  {Anisimov}, \citenamefont {Solovyev}, \citenamefont {Korotin}, \citenamefont
  {Czyzyk},\ and\ \citenamefont {Sawatzky}}]{Anisimov1993}%
  \BibitemOpen
  \bibfield  {author} {\bibinfo {author} {\bibfnamefont {V.~I.}\ \bibnamefont
  {Anisimov}}, \bibinfo {author} {\bibfnamefont {I.~V.}\ \bibnamefont
  {Solovyev}}, \bibinfo {author} {\bibfnamefont {M.~A.}\ \bibnamefont
  {Korotin}}, \bibinfo {author} {\bibfnamefont {M.~T.}\ \bibnamefont {Czyzyk}},
  \ and\ \bibinfo {author} {\bibfnamefont {G.~A.}\ \bibnamefont {Sawatzky}},\
  }\href {http://link.aps.org/doi/10.1103/PhysRevB.48.16929} {\bibfield
  {journal} {\bibinfo  {journal} {Phys. Rev. B}\ }\textbf {\bibinfo {volume}
  {48}},\ \bibinfo {pages} {16929} (\bibinfo {year} {1993})}\BibitemShut
  {NoStop}%
\bibitem [{\citenamefont {Geisler}\ \emph {et~al.}(2017)\citenamefont
  {Geisler}, \citenamefont {Blanca-Romero},\ and\ \citenamefont
  {Pentcheva}}]{Geisler2017}%
  \BibitemOpen
  \bibfield  {author} {\bibinfo {author} {\bibfnamefont {B.}~\bibnamefont
  {Geisler}}, \bibinfo {author} {\bibfnamefont {A.}~\bibnamefont
  {Blanca-Romero}}, \ and\ \bibinfo {author} {\bibfnamefont {R.}~\bibnamefont
  {Pentcheva}},\ }\href {http://link.aps.org/doi/10.1103/PhysRevB.95.125301}
  {\bibfield  {journal} {\bibinfo  {journal} {Phys. Rev. B}\ }\textbf {\bibinfo
  {volume} {95}},\ \bibinfo {pages} {125301} (\bibinfo {year}
  {2017})}\BibitemShut {NoStop}%
\bibitem [{\citenamefont {Grioni}\ \emph {et~al.}(1989)\citenamefont {Grioni},
  \citenamefont {Goedkoop}, \citenamefont {Schoorl}, \citenamefont {de~Groot},
  \citenamefont {Fuggle}, \citenamefont {Sch\"afers}, \citenamefont {Koch},
  \citenamefont {Rossi}, \citenamefont {Esteva},\ and\ \citenamefont
  {Karnatak}}]{Grioni1989}%
  \BibitemOpen
  \bibfield  {author} {\bibinfo {author} {\bibfnamefont {M.}~\bibnamefont
  {Grioni}}, \bibinfo {author} {\bibfnamefont {J.~B.}\ \bibnamefont
  {Goedkoop}}, \bibinfo {author} {\bibfnamefont {R.}~\bibnamefont {Schoorl}},
  \bibinfo {author} {\bibfnamefont {F.~M.~F.}\ \bibnamefont {de~Groot}},
  \bibinfo {author} {\bibfnamefont {J.~C.}\ \bibnamefont {Fuggle}}, \bibinfo
  {author} {\bibfnamefont {F.}~\bibnamefont {Sch\"afers}}, \bibinfo {author}
  {\bibfnamefont {E.~E.}\ \bibnamefont {Koch}}, \bibinfo {author}
  {\bibfnamefont {G.}~\bibnamefont {Rossi}}, \bibinfo {author} {\bibfnamefont
  {J.-M.}\ \bibnamefont {Esteva}}, \ and\ \bibinfo {author} {\bibfnamefont
  {R.~C.}\ \bibnamefont {Karnatak}},\ }\href {\doibase
  10.1103/PhysRevB.39.1541} {\bibfield  {journal} {\bibinfo  {journal} {Phys.
  Rev. B}\ }\textbf {\bibinfo {volume} {39}},\ \bibinfo {pages} {1541}
  (\bibinfo {year} {1989})}\BibitemShut {NoStop}%
\bibitem [{\citenamefont {N\'u\~nez Regueiro}\ \emph
  {et~al.}(1995)\citenamefont {N\'u\~nez Regueiro}, \citenamefont {Altarelli},\
  and\ \citenamefont {Chen}}]{Nunez1995}%
  \BibitemOpen
  \bibfield  {author} {\bibinfo {author} {\bibfnamefont {M.~D.}\ \bibnamefont
  {N\'u\~nez Regueiro}}, \bibinfo {author} {\bibfnamefont {M.}~\bibnamefont
  {Altarelli}}, \ and\ \bibinfo {author} {\bibfnamefont {C.~T.}\ \bibnamefont
  {Chen}},\ }\href {\doibase 10.1103/PhysRevB.51.629} {\bibfield  {journal}
  {\bibinfo  {journal} {Phys. Rev. B}\ }\textbf {\bibinfo {volume} {51}},\
  \bibinfo {pages} {629} (\bibinfo {year} {1995})}\BibitemShut {NoStop}%
\bibitem [{\citenamefont {van~der Laan}(1994)}]{Laan1994}%
  \BibitemOpen
  \bibfield  {author} {\bibinfo {author} {\bibfnamefont {G.}~\bibnamefont
  {van~der Laan}},\ }\href {\doibase 10.1143/JPSJ.63.2393} {\bibfield
  {journal} {\bibinfo  {journal} {Journal of the Physical Society of Japan}\
  }\textbf {\bibinfo {volume} {63}},\ \bibinfo {pages} {2393} (\bibinfo {year}
  {1994})}\BibitemShut {NoStop}%
\bibitem [{\citenamefont {Freeland}\ \emph {et~al.}(2011)\citenamefont
  {Freeland}, \citenamefont {Liu}, \citenamefont {Kareev}, \citenamefont
  {Gray}, \citenamefont {Kim}, \citenamefont {Ryan}, \citenamefont
  {Pentcheva},\ and\ \citenamefont {Chakhalian}}]{Freeland2011}%
  \BibitemOpen
  \bibfield  {author} {\bibinfo {author} {\bibfnamefont {J.~W.}\ \bibnamefont
  {Freeland}}, \bibinfo {author} {\bibfnamefont {J.}~\bibnamefont {Liu}},
  \bibinfo {author} {\bibfnamefont {M.}~\bibnamefont {Kareev}}, \bibinfo
  {author} {\bibfnamefont {B.}~\bibnamefont {Gray}}, \bibinfo {author}
  {\bibfnamefont {J.~W.}\ \bibnamefont {Kim}}, \bibinfo {author} {\bibfnamefont
  {P.}~\bibnamefont {Ryan}}, \bibinfo {author} {\bibfnamefont {R.}~\bibnamefont
  {Pentcheva}}, \ and\ \bibinfo {author} {\bibfnamefont {J.}~\bibnamefont
  {Chakhalian}},\ }\href {http://stacks.iop.org/0295-5075/96/i=5/a=57004}
  {\bibfield  {journal} {\bibinfo  {journal} {EPL (Europhysics Letters)}\
  }\textbf {\bibinfo {volume} {96}},\ \bibinfo {pages} {57004} (\bibinfo {year}
  {2011})}\BibitemShut {NoStop}%
\bibitem [{\citenamefont {Blanca-Romero}\ and\ \citenamefont
  {Pentcheva}(2011)}]{Blanca2011}%
  \BibitemOpen
  \bibfield  {author} {\bibinfo {author} {\bibfnamefont {A.}~\bibnamefont
  {Blanca-Romero}}\ and\ \bibinfo {author} {\bibfnamefont {R.}~\bibnamefont
  {Pentcheva}},\ }\href {http://link.aps.org/doi/10.1103/PhysRevB.84.195450}
  {\bibfield  {journal} {\bibinfo  {journal} {Phys. Rev. B}\ }\textbf {\bibinfo
  {volume} {84}},\ \bibinfo {pages} {195450} (\bibinfo {year}
  {2011})}\BibitemShut {NoStop}%
\bibitem [{\citenamefont {Park}\ \emph {et~al.}(2012)\citenamefont {Park},
  \citenamefont {Millis},\ and\ \citenamefont {Marianetti}}]{Park2012}%
  \BibitemOpen
  \bibfield  {author} {\bibinfo {author} {\bibfnamefont {H.}~\bibnamefont
  {Park}}, \bibinfo {author} {\bibfnamefont {A.~J.}\ \bibnamefont {Millis}}, \
  and\ \bibinfo {author} {\bibfnamefont {C.~A.}\ \bibnamefont {Marianetti}},\
  }\href {http://link.aps.org/doi/10.1103/PhysRevLett.109.156402} {\bibfield
  {journal} {\bibinfo  {journal} {Phys. Rev. Lett.}\ }\textbf {\bibinfo
  {volume} {109}},\ \bibinfo {pages} {156402} (\bibinfo {year}
  {2012})}\BibitemShut {NoStop}%
\bibitem [{\citenamefont {Johnston}\ \emph {et~al.}(2014)\citenamefont
  {Johnston}, \citenamefont {Mukherjee}, \citenamefont {Elfimov}, \citenamefont
  {Berciu},\ and\ \citenamefont {Sawatzky}}]{Johnston2014}%
  \BibitemOpen
  \bibfield  {author} {\bibinfo {author} {\bibfnamefont {S.}~\bibnamefont
  {Johnston}}, \bibinfo {author} {\bibfnamefont {A.}~\bibnamefont {Mukherjee}},
  \bibinfo {author} {\bibfnamefont {I.}~\bibnamefont {Elfimov}}, \bibinfo
  {author} {\bibfnamefont {M.}~\bibnamefont {Berciu}}, \ and\ \bibinfo {author}
  {\bibfnamefont {G.~A.}\ \bibnamefont {Sawatzky}},\ }\href {\doibase
  10.1103/PhysRevLett.112.106404} {\bibfield  {journal} {\bibinfo  {journal}
  {Phys. Rev. Lett.}\ }\textbf {\bibinfo {volume} {112}},\ \bibinfo {pages}
  {106404} (\bibinfo {year} {2014})}\BibitemShut {NoStop}%
\bibitem [{\citenamefont {Zhang}\ \emph {et~al.}(2017)\citenamefont {Zhang},
  \citenamefont {Botana}, \citenamefont {Freeland}, \citenamefont {Phelan},
  \citenamefont {Zheng}, \citenamefont {Pardo}, \citenamefont {Norman},\ and\
  \citenamefont {Mitchell}}]{Zhang2017}%
  \BibitemOpen
  \bibfield  {author} {\bibinfo {author} {\bibfnamefont {J.}~\bibnamefont
  {Zhang}}, \bibinfo {author} {\bibfnamefont {A.~S.}\ \bibnamefont {Botana}},
  \bibinfo {author} {\bibfnamefont {J.~W.}\ \bibnamefont {Freeland}}, \bibinfo
  {author} {\bibfnamefont {D.}~\bibnamefont {Phelan}}, \bibinfo {author}
  {\bibfnamefont {H.}~\bibnamefont {Zheng}}, \bibinfo {author} {\bibfnamefont
  {V.}~\bibnamefont {Pardo}}, \bibinfo {author} {\bibfnamefont {M.~R.}\
  \bibnamefont {Norman}}, \ and\ \bibinfo {author} {\bibfnamefont {J.~F.}\
  \bibnamefont {Mitchell}},\ }\href {http://dx.doi.org/10.1038/nphys4149}
  {\bibfield  {journal} {\bibinfo  {journal} {Nat Phys}\ }\textbf {\bibinfo
  {volume} {advance online publication}} (\bibinfo {year} {2017})}\BibitemShut
  {NoStop}%
\bibitem [{\citenamefont {Wu}\ \emph {et~al.}(2013)\citenamefont {Wu},
  \citenamefont {Soh},\ and\ \citenamefont {Skinner}}]{WuKT2013}%
  \BibitemOpen
  \bibfield  {author} {\bibinfo {author} {\bibfnamefont {K.-T.}\ \bibnamefont
  {Wu}}, \bibinfo {author} {\bibfnamefont {Y.-A.}\ \bibnamefont {Soh}}, \ and\
  \bibinfo {author} {\bibfnamefont {S.~J.}\ \bibnamefont {Skinner}},\ }\href
  {\doibase 10.1016/j.materresbull.2013.05.093} {\bibfield  {journal} {\bibinfo
   {journal} {Materials Research Bulletin}\ }\textbf {\bibinfo {volume} {48}},\
  \bibinfo {pages} {3783} (\bibinfo {year} {2013})}\BibitemShut {NoStop}%
\bibitem [{\citenamefont {Lee}\ \emph {et~al.}(2014)\citenamefont {Lee},
  \citenamefont {Luo}, \citenamefont {Tung}, \citenamefont {Chang},
  \citenamefont {Luo}, \citenamefont {Malshe}, \citenamefont {Gadre},
  \citenamefont {Bhattacharya}, \citenamefont {Nakhmanson}, \citenamefont
  {Eastman}, \citenamefont {Hong}, \citenamefont {Jellinek}, \citenamefont
  {Morgan}, \citenamefont {Fong},\ and\ \citenamefont {Freeland}}]{Lee2014}%
  \BibitemOpen
  \bibfield  {author} {\bibinfo {author} {\bibfnamefont {J.~H.}\ \bibnamefont
  {Lee}}, \bibinfo {author} {\bibfnamefont {G.}~\bibnamefont {Luo}}, \bibinfo
  {author} {\bibfnamefont {I.~C.}\ \bibnamefont {Tung}}, \bibinfo {author}
  {\bibfnamefont {S.~H.}\ \bibnamefont {Chang}}, \bibinfo {author}
  {\bibfnamefont {Z.}~\bibnamefont {Luo}}, \bibinfo {author} {\bibfnamefont
  {M.}~\bibnamefont {Malshe}}, \bibinfo {author} {\bibfnamefont
  {M.}~\bibnamefont {Gadre}}, \bibinfo {author} {\bibfnamefont
  {A.}~\bibnamefont {Bhattacharya}}, \bibinfo {author} {\bibfnamefont {S.~M.}\
  \bibnamefont {Nakhmanson}}, \bibinfo {author} {\bibfnamefont {J.~A.}\
  \bibnamefont {Eastman}}, \bibinfo {author} {\bibfnamefont {H.}~\bibnamefont
  {Hong}}, \bibinfo {author} {\bibfnamefont {J.}~\bibnamefont {Jellinek}},
  \bibinfo {author} {\bibfnamefont {D.}~\bibnamefont {Morgan}}, \bibinfo
  {author} {\bibfnamefont {D.~D.}\ \bibnamefont {Fong}}, \ and\ \bibinfo
  {author} {\bibfnamefont {J.~W.}\ \bibnamefont {Freeland}},\ }\href
  {http://dx.doi.org/10.1038/nmat4039} {\bibfield  {journal} {\bibinfo
  {journal} {Nat Mater}\ }\textbf {\bibinfo {volume} {13}},\ \bibinfo {pages}
  {879} (\bibinfo {year} {2014})}\BibitemShut {NoStop}%
\bibitem [{\citenamefont {Wrobel}(2016)}]{Wrobel2016}%
  \BibitemOpen
  \bibfield  {author} {\bibinfo {author} {\bibfnamefont {F.}~\bibnamefont
  {Wrobel}},\ }\emph {\bibinfo {title} {Structural and electronic properties of
  nickelate heterostructures}},\ \href {http://dx.doi.org/10.18419/opus-8924}
  {\bibinfo {type} {Thesis}},\ \bibinfo  {school} {University of Stuttgart}
  (\bibinfo {year} {2016})\BibitemShut {NoStop}%
\bibitem [{\citenamefont {Lei}\ \emph {et~al.}(2017)\citenamefont {Lei},
  \citenamefont {Golalikhani}, \citenamefont {Davidson}, \citenamefont {Liu},
  \citenamefont {Schlom}, \citenamefont {Qiao}, \citenamefont {Zhu},
  \citenamefont {Chandrasena}, \citenamefont {Yang}, \citenamefont {Gray},
  \citenamefont {Arenholz}, \citenamefont {Farrar}, \citenamefont {Tenne},
  \citenamefont {Hu}, \citenamefont {Guo}, \citenamefont {Singh},\ and\
  \citenamefont {Xi}}]{Lei2017}%
  \BibitemOpen
  \bibfield  {author} {\bibinfo {author} {\bibfnamefont {Q.}~\bibnamefont
  {Lei}}, \bibinfo {author} {\bibfnamefont {M.}~\bibnamefont {Golalikhani}},
  \bibinfo {author} {\bibfnamefont {B.~A.}\ \bibnamefont {Davidson}}, \bibinfo
  {author} {\bibfnamefont {G.}~\bibnamefont {Liu}}, \bibinfo {author}
  {\bibfnamefont {D.~G.}\ \bibnamefont {Schlom}}, \bibinfo {author}
  {\bibfnamefont {Q.}~\bibnamefont {Qiao}}, \bibinfo {author} {\bibfnamefont
  {Y.}~\bibnamefont {Zhu}}, \bibinfo {author} {\bibfnamefont {R.~U.}\
  \bibnamefont {Chandrasena}}, \bibinfo {author} {\bibfnamefont
  {W.}~\bibnamefont {Yang}}, \bibinfo {author} {\bibfnamefont {A.~X.}\
  \bibnamefont {Gray}}, \bibinfo {author} {\bibfnamefont {E.}~\bibnamefont
  {Arenholz}}, \bibinfo {author} {\bibfnamefont {A.~K.}\ \bibnamefont
  {Farrar}}, \bibinfo {author} {\bibfnamefont {D.~A.}\ \bibnamefont {Tenne}},
  \bibinfo {author} {\bibfnamefont {M.}~\bibnamefont {Hu}}, \bibinfo {author}
  {\bibfnamefont {J.}~\bibnamefont {Guo}}, \bibinfo {author} {\bibfnamefont
  {R.~K.}\ \bibnamefont {Singh}}, \ and\ \bibinfo {author} {\bibfnamefont
  {X.}~\bibnamefont {Xi}},\ }\href {\doibase 10.1038/s41535-017-0015-x}
  {\bibfield  {journal} {\bibinfo  {journal} {npj Quantum Materials}\ }\textbf
  {\bibinfo {volume} {2}},\ \bibinfo {pages} {10} (\bibinfo {year}
  {2017})}\BibitemShut {NoStop}%
\end{thebibliography}

\end{document}